\documentclass[11pt]{article}

\def\mytitle#1{\setcounter{equation}{0}
\setcounter{footnote}{0}
\begin{flushleft}\Large\textbf{#1}\end{flushleft}
\vspace{0.25cm}}
\def\myname#1{\leftline{{\large #1}}\vspace{-0.13cm}}
\def\myplace#1#2{\small\begin{flushleft}\textit{#1}\\
\texttt{#2}\end{flushleft}}

\usepackage{graphicx}
\begin{document}

\mytitle{Loop Quantum effects on \textit{Om}-diagnostic and its
Cosmological Implications}

\vskip0.2cm \myname{Prabir Rudra\footnote{prudra.math@gmail.com}}
\myplace{Department of Mathematics, Indian Institute of
Engineering Science and Technology, Shibpur, Howrah-711 103, India.\\
Department of Mathematics, Pailan College of Management and
Technology, Bengal Pailan Park, Kolkata-700 104, India.}{}

\begin{abstract}
In this paper we study the Loop quantum effects on the \textit{Om}
diagnostic and subsequently on the universe. We reconstruct the
\textit{Om} diagnostic in the background of Loop quantum gravity
and then study the behaviour of various Chaplygin gas dark energy
models using the modified diagnostic in a comparative scenario.
The trajectories discriminate the various dark energy models from
each other both in the Einstein gravity as well as Loop quantum
gravity. The Loop quantum effects are also clearly noticeable from
the trajectories in past, present and future universe. We see that
the Loop quantum deviations are highly pronounced in the early
universe, but alleviates as we tend towards the present universe
and continue to decay in future. Thus it puts a big question on
the effectiveness and consequently the suitability of loop quantum
cosmology to explain the future universe.
\end{abstract}

\vspace{5mm}

{\it Pacs. No.: 98.80.-k, 95.36.+x, 98.80.Cq}

\section{Introduction}
With the discovery of the accelerated expansion of the universe
\cite{Perlmutter1, Spergel1}, the concept of dark energy (DE)
\cite{Riess1} have gained prime status in cosmology during the
last decade. Numerous models of DE have been proposed over the
years in order to justify the recent cosmic acceleration. Some of
the well-known models being Chaplygin gas, scalar field, phantom,
etc. Here we consider the Chaplygin gas DE models. The oldest
candidate of the Chaplygin gas cosmology was pure Chaplygin gas
(CG) \cite{Kamenshchik1, Gorini1}. Subsequently Generalized
Chaplygin gas (GCG)\cite{Gorini2, Alam1} was constructed. The
equation of state (EoS) for GCG is given by,
\begin{equation}\label{1}
p_{ch}=-\frac{B}{\rho_{ch}^{\alpha}}
\end{equation}
where $\alpha$ is a constant in the range of $0<\alpha\leq 1$
(obviously $\alpha=1$ corresponds to pure Chaplygin gas) and $B$
is a positive constant. Although GCG was a successful candidate in
playing the role of DE, satisfying almost all the solar system
tests, yet it was plagued by certain cosmological problems like
fine tuning and cosmic coincidence problem. So in the quest of a
better model Modified Chaplygin gas (MCG) \cite{Benaoum1,
Debnath1} came into existence. The MCG EoS is given by,
\begin{equation}\label{2}
p_{m}=A\rho_{m}-\frac{B}{\rho_{m}^\alpha}
\end{equation}
where $0\leq\alpha\leq1$, $A$ and $B$ are positive constants. In
due course MCG was modified into Variable modified Chaplygin gas
(VMCG) and finally into New variable modified Chaplygin gas
(NVMCG). The EoS of VMCG and NVMCG are respectively given as,
\begin{equation}\label{3}
p_{v}=A\rho_{v}-\frac{B(a)}{\rho_{v}^{\alpha}}
\end{equation}
and
\begin{equation}\label{4}
p_{n}=A(a)\rho_{n}-\frac{B(a)}{\rho_{n}^{\alpha}}
\end{equation}
where $A(a)$ and $B(a)$ are positive functions of the cosmological
scale factor $'a'$. In 2003, P. F. Gonz´alez-Diaz \cite{Gonz1}
introduced the generalized cosmic Chaplygin gas (GCCG) model. The
EoS of GCCG is given by,
\begin{equation}\label{5}
p_{g}=-\rho_{g}^{-\alpha}\left[C+\left\{\rho_{g}^{(1+\alpha)}-C\right\}^{-\omega}\right]
\end{equation}
where $C=\frac{A}{1+\omega}-1$, with A being a constant that can
take on both positive and negative values, and $-{\cal
L}<\omega<0$, ${\cal L}$ being a positive definite constant, which
can take on values larger than unity.

In order to complement the statefinder diagnostic \cite{Sahni1}, a
new diagnostic called \textit{Om} \cite{Sahni2}, was proposed in
2008, which helped to distinguish between the energy densities of
various DE models. The \textit{Om} diagnostic is defined as,
\begin{equation}\label{6}
\textit{Om}(z)=\frac{\left(\frac{H(z+1)}{H_{0}}\right)^{2}-1}{\left(z+1\right)^{3}-1}
\end{equation}
Here $H_{0}$ is the present value of Hubble parameter, and $z$ is
the redshift parameter. The advantage of \textit{Om} over the
statefinder parameters is that, \textit{Om} involves only the
first derivative of scale factor, and so it is easier to
reconstruct it from the observational data. For the $\Lambda CDM$
model \textit{Om} diagnostic turns out to be a constant, since
$\Lambda CDM$ is independent of redshift $z$. This is the reason
why we prefer \textit{Om} diagnostic over statefinder parameters
for the present study.

It is believed that general relativity is accurate at small scales
only, and therefore needs modifications at cosmological distances.
Based on the above concept modified gravity evolved as an
alternative to dark energy in order to explain the recent cosmic
acceleration. Here, instead of the matter content of the universe,
the geometry of space-time itself drives the cosmic acceleration.
Some of the well known modified gravity theories are Brane gravity
\cite{Dvali}, Galileon gravity \cite{Nicolis}, Brans-Dicke gravity
\cite{Brans}, Horava-Lifshitz gravity \cite{Horava}, etc. Loop
quantum gravity \cite{Rovelli1, Ashtekar1, Ashtekar2, Ashtekar3,
Bojowald1} evolved a strong counterpart to the above theories,
dealing with the quantum effects of the universe. This property
makes the theory unique and very interesting, since our ultimate
goal is to find a unified theory of general relativity and quantum
mechanics.

In recent years Loop quantum cosmology (LQC) has evolved as a
major candidate for modified gravity models consistent with recent
observational data. It is a non-perturbative and background
independent theory trying to describe the quantum effects of the
universe \cite{Rovelli1, Ashtekar1, Ashtekar2, Ashtekar3,
Bojowald1}. Here a discrete quantum theory replaces the classical
space-time continuum of General Relativity and hence this theory
is identified as a major effort to unify Einstein's General
Relativity with Planck's Quantum theory. Due to this LQC has
attained a prime status in modern cosmology. Extensive research
has been carried out over the past few years in order to develop
the theory and remove the possible shortcomings.

Our motivation is to study the Loop quantum effects on \textit{Om}
diagnostic and subsequently on the universe. It is obvious that we
will need to employ dark energy models for the study. We will
reconstruct the \textit{Om} diagnostic in the background of Loop
quantum gravity (LQG) and then study the behaviour of various
Chaplygin gas DE models using the modified diagnostic. The
trajectories will discriminate the various DE models from each
other. The Loop quantum effects on the universe will be realized
in a comparative scenario. The paper is organized as follows: In
section 2, we study the loop quantum effects on \textit{Om}
diagnostic for different DE models. In section 3, the plots are
analyzed and their cosmological implications are discussed.
Finally the paper ends with a short conclusion in section 4.

\section{Loop quantum effects on the \textit{Om} diagnostic for various Chaplygin gas dark energy models}
Einstein's equations for flat homogeneous and isotropic universe
are given by,
\begin{equation}\label{7}
H^{2}=\frac{\kappa^{2}}{3}\rho
\end{equation}
and
\begin{equation}\label{8}
\dot{H}=-\frac{\kappa^{2}}{2}\left(\rho+p\right)
\end{equation}

The modified Friedmann equations for Loop Quantum Cosmology is
given by \cite{Chen1,Fu1}
\begin{equation}\label{9}
H^2=\frac{\rho}{3}\left(1-\frac{\rho}{\rho_{1}}\right)
\end{equation}
and
\begin{equation}\label{10}
\dot{H}=-\frac{1}{2}\left(\rho+p\right)\left(1-2\frac{\rho}{\rho_{1}}\right)
\end{equation}
Here $\rho$ is the matter density, $p$ is the pressure and
$\rho_{1}=\sqrt{3}\pi^{2}\gamma^{3}G^{2}\hbar$ is the critical
loop quantum density. With the inclusion of this term, the
universe bounces quantum mechanically as the matter energy density
reaches the level of $\rho_{1}$(order of Plank density). Here
$\gamma$ is the dimensionless Barbero-Immirzi parameter.

Using the above field equations in the eqn. (\ref{6}), we get the
\textit{Om} diagnostic in Einstein gravity as given below,
\begin{equation}\label{11}
\textit{Om}_{E}(z)=\frac{-1+\frac{\left(\frac{\kappa
\sqrt{\rho_{E}+\rho_{M}}}{\sqrt{3}}+\frac{\sqrt{3}\kappa
\left(\rho_{E}+\rho_{M}+\rho_{E}\omega\right)}{2\left(1+z\right)
\sqrt{\rho_{E}+\rho_{M}}}\right)^{2}}{H_{0}^{2}}}{-1+(1+z)^3}
\end{equation}
where $\rho=\rho_{E}+\rho_{M}$ is the sum of the energy densities
for dark energy and dark matter. $\omega=\frac{p_{E}}{\rho_{E}}$
is the Equation of State (EoS) parameter of DE.

Using equations (\ref{7}) and (\ref{8}) in equation (\ref{6}), we
get the \textit{Om} diagnostic in Loop quantum gravity is given
by,
\begin{equation}\label{12}
\textit{Om}_{L}(z)=\frac{-1+\frac{\left(\frac{\kappa
\sqrt{(\rho_{E}+\rho_{M})
\left(1-\frac{\rho_{E}+\rho_{M}}{\rho_{1}}\right)}}{\sqrt{3}}+\frac{\sqrt{3}\kappa\left(1-\frac{2
(\rho_{E}+\rho_{M})}{\rho_{1}}\right)
(\rho_{E}+\rho_{M}+\omega\rho_{E})}{2(1+z)
\sqrt{(\rho_{E}+\rho_{M})
\left(1-\frac{\rho_{E}+\rho_{M}}{\rho_{1}}\right)}}\right)^2}{H_{0}^2}}{-1+(1+z)^3}
\end{equation}

\subsection{Generalized Chaplygin gas}
From the EoS of GCG we get the expression for the energy density
of GCG as,
\begin{equation}\label{13}
\rho_{ch}=\left(B+C_{1}\left(1+z\right)^{3(1+\alpha)}\right)^{\frac{1}{1+\alpha}}
\end{equation}
where $C_{1}$ is the constant of integration. Using equation
(\ref{13}) in equations (\ref{11}) and (\ref{12}) we get the
expressions for \textit{Om} diagnostic in Einstein gravity and
Loop quantum gravity respectively as,
\begin{equation}\label{14}
\textit{Om}_{E}^{ch}=\frac{1}{\left(-1+(1+z)^3\right)}\left[-1+\frac{1}{H_{0}^2}\left(\frac{\kappa
\sqrt{\rho_{ch}+\rho_{M}}}{\sqrt{3}}+\frac{\sqrt{3}\kappa
\left(\rho_{ch}+\rho_{M}+\omega_{ch}\rho_{ch}\right)}{2(1+z)
\sqrt{\rho_{ch}+\rho_{M}}}\right)^2\right]
\end{equation}
and
\begin{equation}\label{15}
\textit{Om}_{L}^{ch}=\frac{1}{-1+(1+z)^3}\left[-1+\frac{1}{H_{0}^2}
\left(\frac{\kappa\sqrt{\left(\rho_{ch}+\rho_{M}\right)
\left(1-\frac{\rho_{ch}+\rho_{M}}{\rho_{1}}\right)}}{\sqrt{3}}
+\frac{\left(\sqrt{3}\kappa\left(1-\frac{2\left(\rho_{ch}+\rho_{M}\right)}{\rho_{1}}\right)
\left(\rho_{ch}+\rho_{M}+\rho_{ch}\omega_{ch}
\right)\right)}{2(1+z) \sqrt{\left(\rho_{ch}+\rho_{M}\right)
\left(1-\frac{\rho_{ch}+\rho_{M}}{\rho_{1}}\right)}}\right)^{2}\right]
\end{equation}
where $\omega_{ch}=\frac{p_{ch}}{\rho_{ch}}$

\vspace{2mm}
\begin{figure}
~~~~~~~~~~~~~~~~~\includegraphics[height=3in]{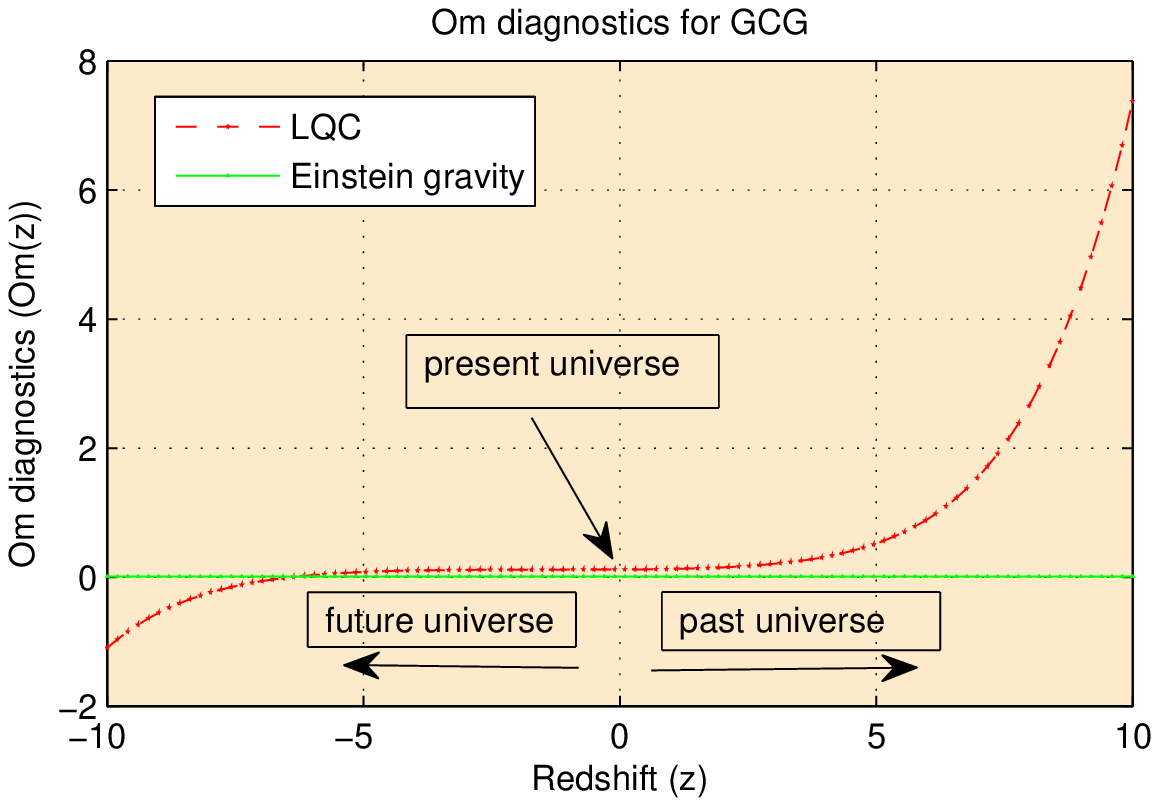}~~~~\\
\vspace{5mm}
~~~~~~~~~~~~~~~~~~~~~~~~~~~~~~~~~~~~~~~~~~~~~~~~~~~~~~Fig.1~~~~~~~~~~~~~~~~~~~~~~~~~~~~~~~~~\\
\vspace{1mm} \textsl{Fig 1 : trajectories in Einstein gravity and
LQC for GCG. The other parameters are considered as $B=1.5,
\alpha=0.5, \kappa=1,
\rho_{1}=0.2, H_{0}=72, \omega_{ch}=-1/3, C_{1}=1.$}\\

\end{figure}

\subsection{Modified Chaplygin gas}
From the EoS of MCG we get the expression for energy density as,
\begin{equation}\label{16}
\rho_{mcg}=\left[\frac{B}{1+A}+C_{2}\left(1+z\right)^{3(1+A)(1+\alpha)}
\right]^{\frac{1}{1+\alpha}}
\end{equation}
where $C_{2}$ is the integration constant. Using eqn. (\ref{16})
is eqns. (\ref{11}) and (\ref{12}) we get the \textit{Om}
diagnostic for MCG in Einstein gravity and LQC respectively as
below,
\begin{equation}\label{17}
\textit{Om}_{E}^{m}=\frac{1}{-1+(1+z)^3}\left[-1+\frac{1}{H_{0}^2}\left(\frac{\kappa
\sqrt{\rho_{mcg}+\rho_{M}}}{\sqrt{3}}+\frac{\left(\sqrt{3}\kappa
\left(\rho_{mcg}+\rho_{M}+\rho_{mcg}\omega_{m}\right)\right)}
{2(1+z)\sqrt{\rho_{mcg}+\rho_{M}}}\right)^{2}\right]
\end{equation}
and

\begin{equation}\label{18}
\textit{Om}_{L}^{m}=\frac{1}{-1+(1+z)^3}\left[-1+\frac{1}{H_{0}^2}
\left(\frac{\kappa\sqrt{\left(\rho_{mcg}+\rho_{M}\right)
\left(1-\frac{\rho_{mcg}+\rho_{M}}{\rho_{1}}\right)}}{\sqrt{3}}
+\frac{\sqrt{3}\kappa\left(1-\frac{2\left(\rho_{mcg}+\rho_{M}\right)}{\rho_{1}}\right)
\left(\rho_{mcg}+\rho_{M}+\rho_{mcg}\omega_{m}
\right)}{2(1+z)\sqrt{\left(\rho_{mcg}+\rho_{M}\right)
\left(1-\frac{\left(\rho_{mcg}+\rho_{M}\right)}{\rho_{1}}\right)}}\right)^2\right]
\end{equation}
where $\omega_{m}=\frac{p_{m}}{\rho_{m}}$

\vspace{2mm}
\begin{figure}
~~~~~~~~~~~~~~~~~\includegraphics[height=3in]{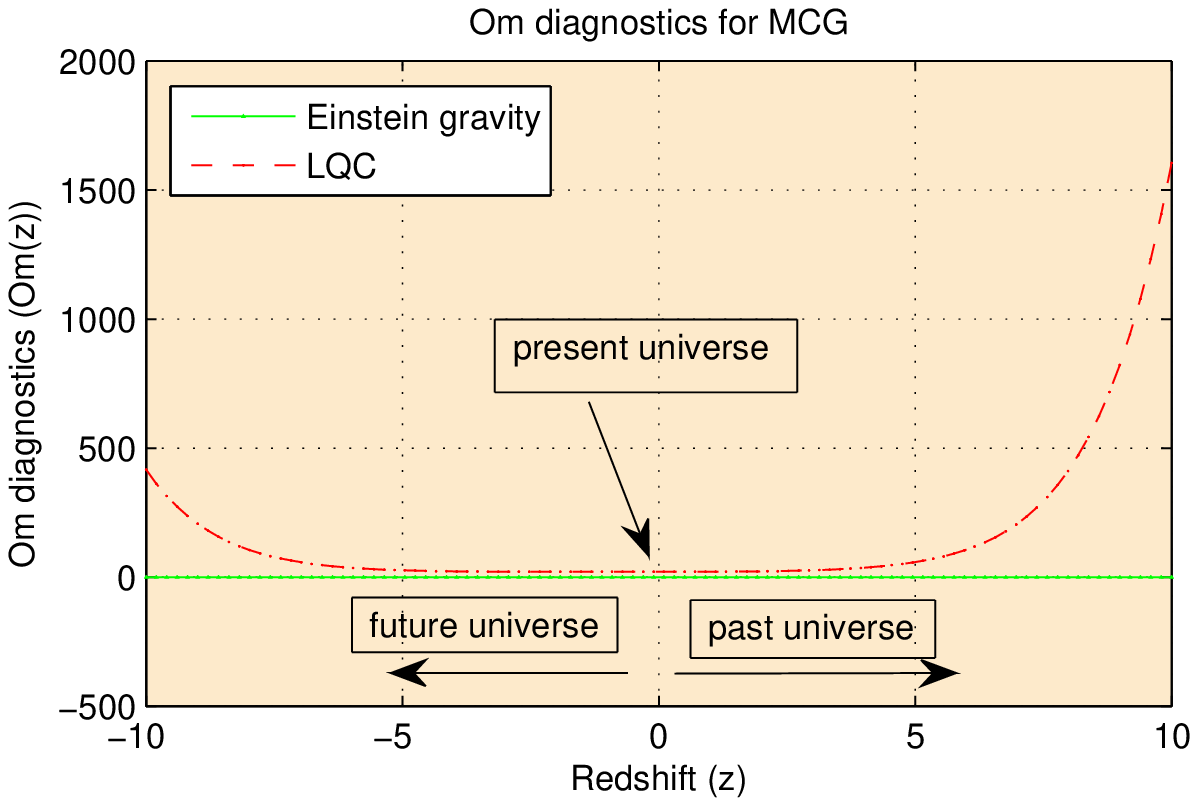}~~~~\\
\vspace{5mm}
~~~~~~~~~~~~~~~~~~~~~~~~~~~~~~~~~~~~~~~~~~~~~~~~~~~~~~~Fig.2~~~~~~~~~~~~~~~~~~~~~~~~~~~~~~~~~\\
\vspace{1mm} \textsl{Fig 2 : trajectories in Einstein gravity and
LQC for MCG. The other parameters are considered as $B=1.5, A=1/3,
\alpha=0.5, \kappa=1,
\rho_{1}=0.2, H_{0}=72, \omega_{m}=-1/3, C_{2}=1.$}\\
\end{figure}

\subsection{Variable Modified Chaplygin gas}
The EoS of VMCG is given by eqn. (\ref{3}). We consider
$B(a)=B_{0}a^{-m}$, where $B_{0}$ and $m$ are constants. Using the
EoS of VMCG, we get the expression for energy density as,
\begin{equation}\label{19}
\rho_{v}=\left[\frac{B_{0}\left(1+\alpha\right)\left(1+z\right)^{m}}{\left(A+1\right)\left(\alpha+1\right)-m}+C_{3}\left(1+z\right)^{\left(\alpha+1\right)\left(A+1\right)}\right]^{\frac{1}{1+\alpha}}
\end{equation}
where $C_{3}$ is the constant of integration. Now using the above
equation in eqns. (\ref{11}) and (\ref{12}), we get the
expressions of \textit{Om} diagnostic for VMCG in Einstein gravity
and LQC respectively as,

\begin{equation}\label{20}
\textit{Om}_{E}^{v}=\frac{1}{-1+(1+z)^3}\left[-1+
\frac{1}{H_{0}^2}\left(\frac{\kappa\sqrt{\rho_{v}+\rho_{M}}}{\sqrt{3}}
+\frac{\left(\sqrt{3}\kappa\left(\rho_{v}+\rho_{M}+\rho_{v}\omega_{v}\right)\right)}
{2(1+z)\sqrt{\rho_{v}+\rho_{M}}}\right)^2\right]
\end{equation}
and

\begin{equation}\label{21}
\textit{Om}_{L}^{v}=\frac{1}{-1+(1+z)^3}\left[-1+\frac{1}{H_{0}^2}\left(\frac{\kappa\sqrt{\left(\rho_{v}+\rho_{M}\right)
\left(1-\frac{\rho_{v}+\rho_{M}}{\rho_{1}}\right)}}{\sqrt{3}}+\frac{\left(\sqrt{3}\kappa\left(1-\frac{2
\left(\rho_{v}+\rho_{M}\right)}{\rho_{1}}\right)\left(\rho_{v}+\rho_{M}+\rho_{v}\omega_{v}\right)\right)}
{2(1+z)\sqrt{\left(\rho_{v}+\rho_{M}\right)
\left(1-\frac{\rho_{v}+\rho_{M}}{\rho_{1}}\right)}}\right)^2\right]
\end{equation}
where $\omega_{v}=\frac{p_{v}}{\rho_{v}}$

\vspace{2mm}
\begin{figure}
~~~~~~~~~~~~~~~~~\includegraphics[height=3in]{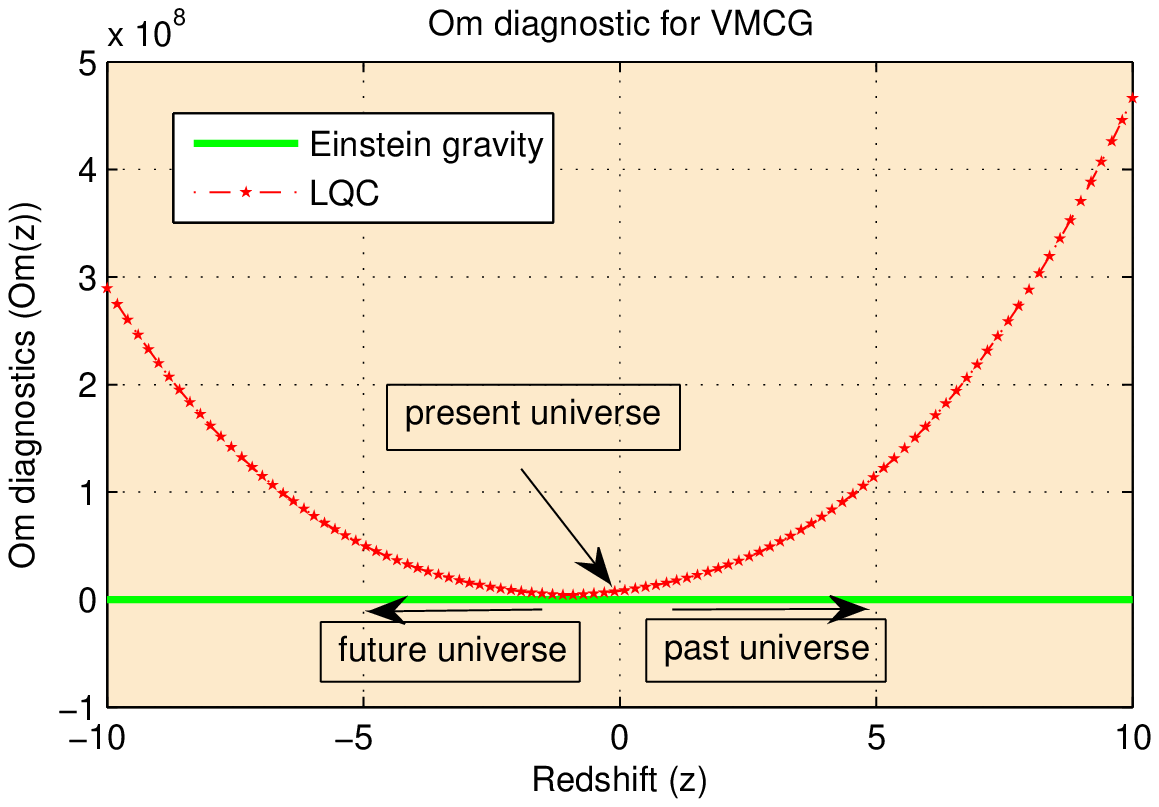}~~~~\\
\vspace{5mm}
~~~~~~~~~~~~~~~~~~~~~~~~~~~~~~~~~~~~~~~~~~~~~~~~~~~~~~~Fig.3~~~~~~~~~~~~~~~~~~~~~~~~~~~~~~~~~\\
\vspace{1mm} \textsl{Fig 3 : trajectories in Einstein gravity and
LQC for VMCG. The other parameters are considered as $A=1/3,
B_{0}=2, \alpha=0.5, \kappa=1,
\rho_{1}=10^{-7}, H_{0}=72, \omega_{v}=-1/3, C_{3}=1, m=4.$}\\
\end{figure}

\subsection{New Variable Modified Chaplygin gas}
We consider $A(a)=A_{0}a^{-n}$ and $B(a)=B_{0}a^{-m}$ (where
$A_{0}$, $B_{0}$, $m$ and $n$ are constants) in the EoS of NVMCG
(\ref{4}) and get the expression for energy density as,
\begin{equation}\label{22}
\rho_{n}=\left(1+z\right)^{3}\exp\left(\frac{3A_{0}}{n\left(1+z\right)^{n}}\right)\left[C_{4}
+\frac{B_{0}}{A_{0}}\left(\frac{3A_{0}\left(1+\alpha\right)}{n}\right)^{\frac{3(1+\alpha)+n-m}{n}}\times
\Gamma\left(\frac{m-3\left(1+\alpha\right)}{n},
\frac{3A_{0}\left(1+\alpha\right)}{n\left(1+z\right)^{n}}\right)\right]^{\frac{1}{1+\alpha}}
\end{equation}
where $C_{4}$ is the integration constant and $\Gamma (s,t)$ is
the upper incomplete gamma function. Using the above equation in
eqns. (\ref{11}) and (\ref{12}), we get the following expressions
for \textit{Om} diagnostic in Einstein gravity and LQC
respectively,

\begin{equation}\label{23}
\textit{Om}_{E}^{n}=\frac{1}{-1+(1+z)^3}\left[-1+\frac{1}{H_{0}^2}\left(\frac{\kappa
\sqrt{\rho_{M}+\rho_{n}}}{\sqrt{3}}+\frac{\left(\sqrt{3}\kappa
\left(\rho_{M}+\rho_{n}+\omega_{n}\rho_{n}\right)\right)}{2(1+z)
\sqrt{\rho_{M}+\rho_{n}}}\right)^2\right]
\end{equation}
and
\begin{equation}\label{24}
\textit{Om}_{L}^{n}=\frac{1}{-1+(1+z)^3}\left[-1+
\frac{1}{H_{0}^2}\left(\frac{\sqrt{3}\kappa\left(\rho_{M}+\rho_{n}+\omega_{n}\rho_{n}\right)
\left(1-\frac{2 \left(\rho_{M}+\rho_{n}\right)}{\rho_{1}}\right)}
{2(1+z)\sqrt{\left(\rho_{M}+\rho_{n}\right)
\left(1-\frac{\rho_{M}+\rho_{n}}{\rho_{1}}\right)}}+\frac{\kappa}{\sqrt{3}}\sqrt{
\left(\left(\rho_{M}+\rho_{n}\right)
\left(1-\frac{\rho_{M}+\rho_{n}}{\rho_{1}}\right)\right)}\right)^2\right]
\end{equation}
where $\omega_{n}=\frac{p_{n}}{\rho_{n}}$

\vspace{2mm}
\begin{figure}
~~~~~~~~~~~~~~~~~\includegraphics[height=3in]{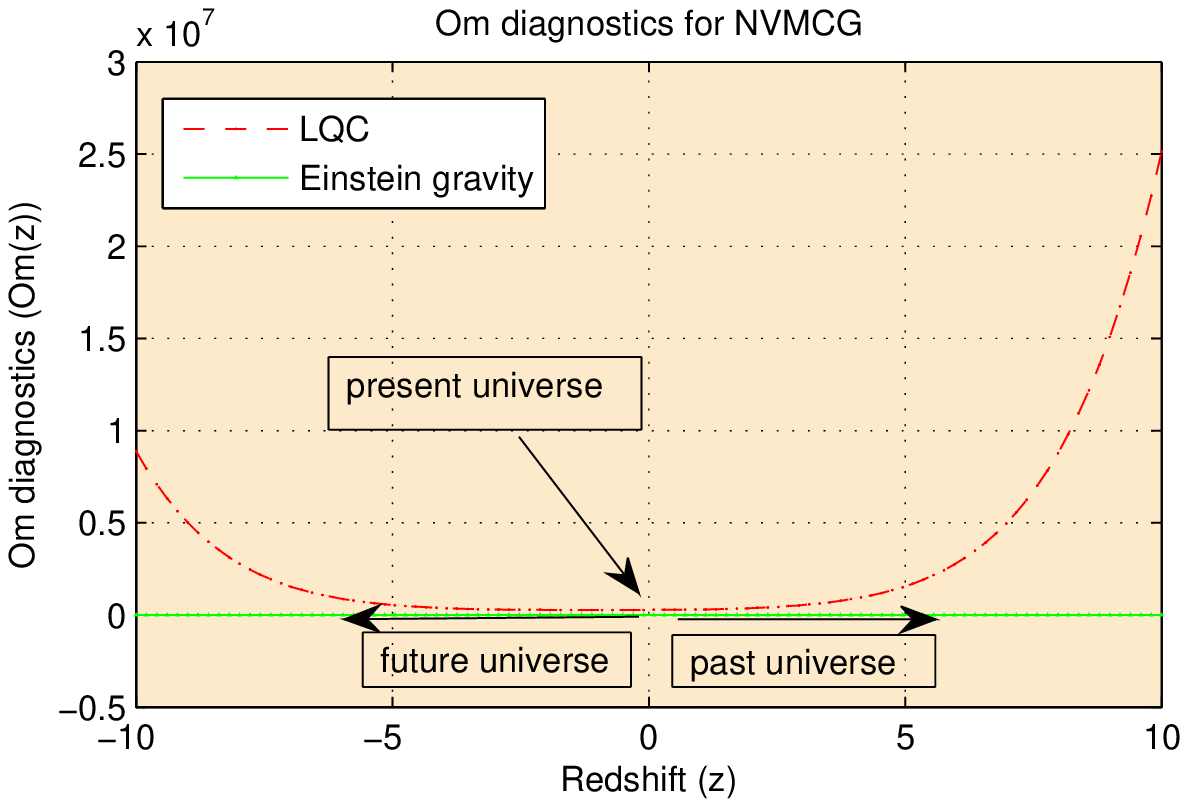}~~~~\\
\vspace{5mm}
~~~~~~~~~~~~~~~~~~~~~~~~~~~~~~~~~~~~~~~~~~~~~~~~~~~~~~~Fig.4~~~~~~~~~~~~~~~~~~~~~~~~~~~~~~~~~\\
\vspace{1mm} \textsl{Fig 4 : trajectories in Einstein gravity and
LQC for NVMCG. The other parameters are considered as $A_{0}=1.2,
B_{0}=2, \alpha=0.5, \kappa=1,
\rho_{1}=0.2, H_{0}=72, \omega_{n}=-1/3, C_{4}=1, m=4, n=2.$}\\
\end{figure}

\subsection{Generalized Cosmic Chaplygin gas}
The energy density of GCCG is given by,
\begin{equation}\label{25}
\rho_{g}=\left[C+\left\{1+C_{5}\left(1+z\right)^{3(1+\alpha)(1+\omega)}\right\}^\frac{1}{1+\omega}\right]^\frac{1}{1+\alpha}
\end{equation}
where $C_{5}$ is the integration constant. Using the above
equation in eqns. (\ref{11}) and (\ref{12}), we get the
expressions for \textit{Om} diagnostic as given below,
\begin{equation}\label{26}
\textit{Om}_{E}^{g}=\frac{1}{-1+(1+z)^3}\left[-1+\frac{1}{H_{0}^2}
\left(\frac{\kappa\sqrt{\rho_{g}+\rho_{M}}}{\sqrt{3}}+\frac{\left(\sqrt{3}\kappa\left(\rho_{g}+\rho_{M}+
\rho_{g}\omega_{g}\right)\right)}{2(1+z)\sqrt{\rho_{g}+\rho_{M}}}\right)^2\right]
\end{equation}
and
\begin{equation}\label{27}
\textit{Om}_{L}^{g}=\frac{1}{-1+(1+z)^3}\left[-1+
\frac{1}{H_{0}^2}\left(\frac{\sqrt{3}\kappa\left(\rho_{M}+\rho_{g}+\omega_{g}\rho_{g}\right)
\left(1-\frac{2 \left(\rho_{M}+\rho_{g}\right)}{\rho_{1}}\right)}
{2(1+z)\sqrt{\left(\rho_{M}+\rho_{g}\right)
\left(1-\frac{\rho_{M}+\rho_{g}}{\rho_{1}}\right)}}+\frac{\kappa}{\sqrt{3}}\sqrt{
\left(\rho_{M}+\rho_{g}\right)
\left(1-\frac{\rho_{M}+\rho_{g}}{\rho_{1}}\right)}\right)^2\right]
\end{equation}
where $\omega_{g}=\frac{p_{g}}{\rho_{g}}$

\vspace{2mm}
\begin{figure}
~~~~~~~~~~~~~~~~~\includegraphics[height=3in]{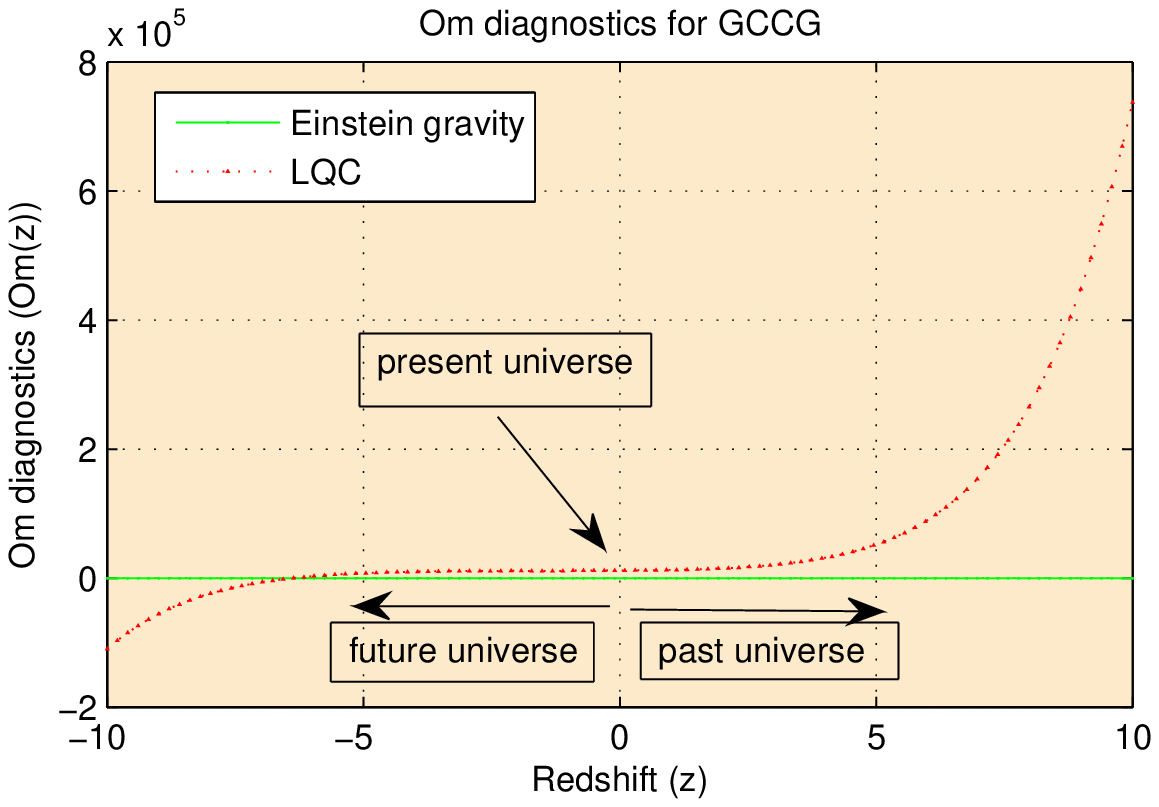}~~~~\\
\vspace{5mm}
~~~~~~~~~~~~~~~~~~~~~~~~~~~~~~~~~~~~~~~~~~~~~~~~~~~~~~~Fig.5~~~~~~~~~~~~~~~~~~~~~~~~~~~~~~~~~\\
\vspace{1mm} \textsl{Fig 5 : trajectories in Einstein gravity and
LQC for GCCG. The other parameters are considered as $C=5,
\alpha=0.5, \kappa=1,
\rho_{1}=0.2, H_{0}=72, \omega_{g}=-1/3, C_{5}=1, \omega=-0.5.$}\\
\end{figure}

\section{Graphical Analysis and Cosmological Implications}
Plots showing the trajectories of \textit{Om} diagnostic for both
Einstein gravity and Loop quantum gravity are given above. The
loop quantum deviations are clearly visible in the figures.
\textit{Fig.1} shows the trajectories for GCG. $\omega$ has been
considered as $-1/3$. So we have considered an accelerating
universe dominated by dark energy. We can see that for early
universe(i.e. for higher redshifts), the loop quantum deviations
are more pronounced. It reduces in magnitude as we come close to
the present epoch. The portion of the plot $z<-1$ is obviously
un-physical, yet it has been retained in the figure in order to
give it a complete shape and for a better understanding of the
\textit{Om} dynamics. It is be seen that the two trajectories meet
at around $z=-6$, which is totally un-physical as far as our
notion of cosmology is concerned (because $z>-1$). Therefore we
can say that the trajectory for loop quantum gravity attains an
asymptotic behaviour around the present regime and continue to do
so in the future universe.

\textit{Fig2} shows the plot of \textit{Om} diagnostics for MCG.
Clear deviations in trajectories are visible, corresponding to the
loop quantum effects. But this differs from the case of GCG in the
sense that the two trajectories never intersect each other. In fig
2, the trajectories for MCG are obtained both for Einstein gravity
and LQC. In this paper, we have actually used the Om- diagnostics,
not only as a tool to differentiate between various DE models but
also we have analogously used the Om-trajectories to get
information about the state of universe at a certain cosmological
time. The states when compared between the Einstein gravity and
LQC at the same cosmological time gives us the Loop quantum
deviations suffered by the universe. From the fig.2 it is visible
that the model suffers large quantum deviations in the early
universe. The deviations alleviate as we reach the present epoch
and they coincide in the present time $(z=0)$ , as is expected. As
we move into the future universe, i.e. $z<0$, we can see that the
deviations again become large and continue to grow as $z$
decreases. So in terms of deviations, the scenario is oscillatory
about the present time $(z=0)$. It is actually an oscillating
scenario between the two large deviation regimes for $z>0$ and
$z<0$, with a no deviation regime at $z=0$. This can probably be
attributed to the barotropic term (first term) in the EoS of MCG.

\textit{Fig3}, represents the plot for VMCG. In contrast to the
previous two plots, this plot shows greater loop quantum
deviations around the present regime. Although the trajectories
coincide around $z=0$, yet the region of coincidence is far
smaller than the previous plots. In fact there is a sharp point of
coincidence around $z=0$ and then it immediately takes off showing
larger deviations compared to the other plots. The reason behind
this is obviously the chosen power law form of the function
$B(a)$. In \textit{Fig4}, plot have been generated for NVMCG. The
plot almost resembles that of MCG. Finally in \textit{Fig5}, we
witness the plot for GCCG, which shows the same trend as that of
GCG.

In almost all the plots the Loop quantum effects are substantial
in the early universe and gets alleviated to a comparable scenario
around the present regime. The early universe was dominated by
mass-less radiation. This lack or absence of mass is primarily
responsible for the large quantum effect. Although radiation
pressure exists which gives rise to momentum, yet the effect is
far lesser than that of matter due to the absence of mass. As the
universe tends towards the present regime, matter content in the
universe increases considerably. Both dark and visible matter come
into co-existence. This automatically alleviates the loop quantum
effect on the universe and so the trajectory almost coincides with
that of Einstein gravity. Moreover, we see that not only in the
present epoch but also it preserves its comparable nature in the
future universe. Obviously the difference between two gravity
theories at present depends on the critical loop quantum density
$\rho_{1}$. $\rho_{1}$ being an adjustable parameter, it is quite
obvious to think that $\rho_{1}$ can be effectively fine tuned in
order to realize different cosmological scenarios. But our
analysis has shown that there is not much alteration in the
scenario, with alteration in the value of $\rho_{1}$ for most of
the DE models. It is only in the case of MCG, that we have to
considerably fine tune the value of $\rho_{1}$ in order to
generate coinciding trajectories for both the gravity theories,
which is fundamental. So in a nutshell, in the present dark energy
dominated epoch the loop quantum effects are diminished to such
extent that it almost coincides with Einstein gravity, which
follows fundamentally from the theory (LQC) itself. But the
interesting thing that comes out from this study is that in the
future regime $(-1<z<0)$ the loop quantum effects continue to
decay showing comparable behaviour to Einstein gravity. In that
case the obvious questions that follow: How effective will LQC be
in the future regime? Will it be able to describe the future
universe as effectively as it has done till now? Any modified
gravity theory evolves as a modification to Einstein gravity. If
after all these, it cannot show considerable deviations from
Einstein gravity, is there any necessity for it? Is it not that
the modifications are not satisfactory. Do we really need that
theory? These are the basic questions that arise from the
behaviour of LQC in the future universe. Although on the basis of
the results obtained from the above theoretical study, we cannot
make a strong statement or reach a conclusion, yet the study does
give us something to think about. Something that may change the
present and the future cosmological scenario. Something that may
rule out Loop quantum cosmology in near future and subsequently
pave the way for alternative theories like string theory gaining
prominence.

\section{Conclusion}

In this study we have basically made an attempt to study the loop
quantum effects on the universe by studying its effects on the
\textit{Om} diagnostics. Modified equations for the diagnostics
were obtained in the background of loop quantum gravity. To study
the effects, trajectories were generated for different class of
Chaplygin gas dark energy models. The study revealed that for
almost all the models the quantum effects are highly pronounced in
the early universe, which was dominated by mass-less radiation
energy. But the more we moved towards the present regime the
scenario completely changed as the trajectories for loop quantum
gravity became almost comparable to those of Einstein gravity,
thus exhibiting a strange alleviation of the quantum effect. Not
only the present epoch, the alleviation continued in the future
rendering the quantum gravity ineffective. Now the question arises
that does the study really opens a gateway to something
substantial in near future? Can loop quantum cosmology be ruled
out for the future universe? Or is it that the theory is not
consistent enough and requires further modifications! Obviously
based on this study, it will not be fair to conclude anything, but
it does give us a hope of something new. For the time being we
keep it an open question, subject to further extensive research.\\

{\bf Acknowledgement:}\\\\
The author registers his whole hearted gratitude to Swarnali
Sharma, for her kind assistance regarding the figures. The author
is also thankful to Dr. Rajesh Kumar Neogy for helpful
discussions. The author sincerely acknowledges the anonymous
referee for his or her constructive comments which helped him to
improve the quality of the manuscript.\\

\end{document}